# Novel porous polymorphs of zinc cyanide with rich thermal and mechanical behavior

Fabien Trousselet,[1] Anne Boutin,[1, *] and François-Xavier Coudert[2, †]

[1]*École Normale Supérieure, PSL Research University, Département de Chimie,*
*Sorbonne Universités – UPMC Univ Paris 06, CNRS UMR 8640 PASTEUR, 24 rue Lhomond, 75005 Paris, France*
[2]*PSL Research University, Chimie ParisTech – CNRS, Institut de Recherche de Chimie Paris, 75005 Paris, France*

We investigate the feasibility of four-connected nets as hypothetical zinc cyanide polymorphs, as well as their thermal and mechanical properties, through quantum chemical calculations and molecular dynamics simulations. We confirm the metastability of the two porous phases recently discovered experimentally [S. H. Lapidus et al., *J. Am. Chem. Soc.* **135**, 7621–7628 (2013)], suggest the existence of 7 novel phases of Zn(CN)$_2$, and show that isotropic negative thermal expansion is a common occurrence among all members of this family of materials, with thermal expansion coefficients close to that of the dense *dia-c* phase. In constrast, we find a wide variety in the mechanical behavior of these porous structures, with framework-dependent anisotropic compressibilities. All porous structures, however, show pressure-induced softening leading to a structural transition at modest pressure.

## I. INTRODUCTION

Molecular framework materials, including the ubiquitous metal-organic frameworks (MOFs), demonstrate a wide variety of behavior in response to changes in temperature as well as the application of external pressure.[1] Among the most unusual and high sought-after reponses are anomalous thermal and mechanical properties, such as negative thermal expansion (NTE),[2,3] negative linear compressibility (NLC),[4,5] pressure-induced softening,[6] and auxeticity (presence of a negative Poisson's ratio in one or more directions).[7,8] In addition to being rather counter-intuitive, these properties have applications in creating zero-thermal expansion composites, artificial muscles, actuators, and blast mitigating materials. In addition, they are of particular interest in nanoporous materials, where their coupling with guest adsorption can be leveraged for sensing applications and in the making of multifunctional materials.

In this context, the prototypical molecular framework zinc cyanide, Zn(CN)$_2$, is a very interesting system. In ambient conditions, it forms a dense interpenetrated structure composed of two diamondoid (*dia*) frameworks, where the tetrahedral Zn$^{2+}$ cations are interconnected by linear Zn–CN–Zn linkages with disordered cyanide anions.[9] This *dia-c* phase of Zn(CN)$_2$ exhibits very strong isotropic NTE, with a volumetric coefficient of thermal expansion of $\alpha_V = -51$ MK$^{-1}$, as well as extreme pressure-induced softening.[10] Moreover, it was recently shown that under compression with certain molecular fluids (water, methanol, ethanol) or fluid mixtures, *dia-c*-Zn(CN)$_2$ can undergo reconstructive transitions to porous non-interpenetrated polymorphs with topologies *dia* and *lon*, which are then retained upon release to ambient pressure.[11]

This entirely novel approach to form new porous phases of framework materials raises however several unanswered questions. In particular, do the porous Zn(CN)$_2$ frameworks share the anomalous thermal and mechanical properties of their parent compound, the interpenetrated framework *dia-c*-Zn(CN)$_2$? What is the extent of the influence of the framework topology on the thermal expansion coefficient, the linear compressibility, and their anisotropy? And finally, are there other experimentally feasible porous frameworks in this family, i.e. other low-density polymorphs of Zn(CN)$_2$ energetically close to the experimentally observed *dia* and *lon* structures?

Here, we select and investigate through quantum chemistry calculations and molecular dynamics simulations a set of frameworks suitable as zinc cyanide polymorphs. First we address the relative stability and structural properties of the corresponding structures at T= 0 K. We then turn to their thermal stability and thermal properties in the 50–500 K range, as well as their behavior under isostatic compression and mechanical properties in general.

## II. SYSTEMS AND METHODS

### A. Selection of frameworks

The use of computational chemistry to study families of polymorphs has long been proven a successful tool for the investigation of structure–property relationships, energetic stability and experimental feasibility. Seminal work in this area dates back to the early computational studies of all-silica zeolite frameworks with classical force fields,[12,13] and has shown continuing development even in recent years, with focus shifting on a larger number of physical properties[14] or quantum chemical calculations.[15] Other families of nanoporous materials have also been studied including the very topical metal–organic frameworks, and in particular the family of Zeolitic Imidazolate Frameworks (or ZIFs).[16,17] This is part of a broader effort at computationally-aided discovery of novel materials, though high-throughput screening of candidate materials based on the tools of theoretical chemistry.[18,19]

For this study, we first generated a set of plausible candidates for porous polymorphs of Zn(CN)$_2$, with diverse framework topologies. To that aim, we used the RCSR online database[20,21] of periodic 3-dimensional nets. We filtered the 2288 nets in the database (at the time of the research) to select those that are structurally close to the known Zn(CN)$_2$ phases, with Zn$^{2+}$ ions (vertices of the net) tetrahedrally coordinated by cyanide anions (edges). Of the 629 four-coordinated nets in the RCSR database, we selected structures with density in the range of 0.5 to 1.5 g.cm$^{-3}$, near-uniform edge lengths, and Zn–Zn–Zn angles close to 109.5° (so that the deformations of the [Zn(CN)$_4$]$^{2-}$ tetrahedra are not too large).

This search resulted in a set of 11 plausible nets, listed in Table I with their nominal density, space group, and number of edges and vertices. This set contains the interpenetrated dense phase *dia-c* (the only interpenetrated net included) as well as the experimentally known porous phases *dia* and *lon*. For each of these nets, we constructed a Zn(CN)₂ structure by placing CN⁻ anions along the edges of the net. This ordering of the CN⁻ anions, which are thought to be disordered in the material, was shown by earlier theoretical work to have negligible influence on the structural and dynamic properties of Zn(CN)₂.[22] We thus chose the orientations of the anions so as to obtain for each framework the highest symmetry possible, i.e. to be in a subgroup of the net space group with maximal symmetry (Table I).

## B. Quantum chemical calculations and molecular dynamics simulations

Once the frameworks chosen, we performed full geometry optimization of the structures using quantum chemistry calculations at the Density Functional Theory (DFT) level, optimizing both atomic positions and unit cell parameters. These DFT calculations were performed using the CRYSTAL14 code[23], with localized atomic basis sets, full accounting for the symmetry operators of the crystal, and the B3LYP hybrid exchange–correlation functional.[24] We used the Grimme ''D2'' dispersion corrective terms[25] to effectively take into account long-range (van der Waals) interactions.

Then, in order to study the influence of temperature and mechanical pressure on the frameworks, we used classical force field-based molecular dynamics simulations in the isostress $(N, \sigma, T)$ ensemble with the DL_POLY software.[26] The force field used is that derived by Fang et al. for Zn(CN)₂.[22] It includes (i) a Morse potential governing the distances between a Zn atom and its C/N coordinating neighbors, (ii) several terms creating an energy cost for C/N-Zn-C/N and Zn-C/N-C/N angles deviating from their values — respectively 109.5° and 180° — expected in an ideal tetrahedral geometry, (iii) Coulomb interactions between effective atom-centered point charges that take into account multipolar effects, and (iv) van der Waals interactions described by a Buckingham-type potential. Importantly, given the strength of the CN bonding, cyanide ions are considered as rigid rods.

For each of the 11 frameworks, a supercell of the crystallographic unit cell obtained by quantum chemistry calculations was selected, so that simulation box sizes were close to 40 Å in each direction (in all cases between 30 and 50 Å, with aspect ratios as close to 1 as possible). The simulations were performed using the DL_POLY software[26] in the $(N, \sigma, T)$ ensemble, using a Nosé-Hoover algorithm[27] allowing anisotropic deformations of the unit cell. Thermostat and barostat relaxation times were set to 1 ps. Equations of motion were integrated using a leapfrog algorithm, with timestep of 1 fs.[28] For long-range interactions, the Ewald summation method was used with a precision of 10⁻⁶. Unless specified, the time of each simulation is 200 ps, including 20 ps of equilibration. For calculation of elastic stiffness tensors, total simulations times were upped to 5 ns.

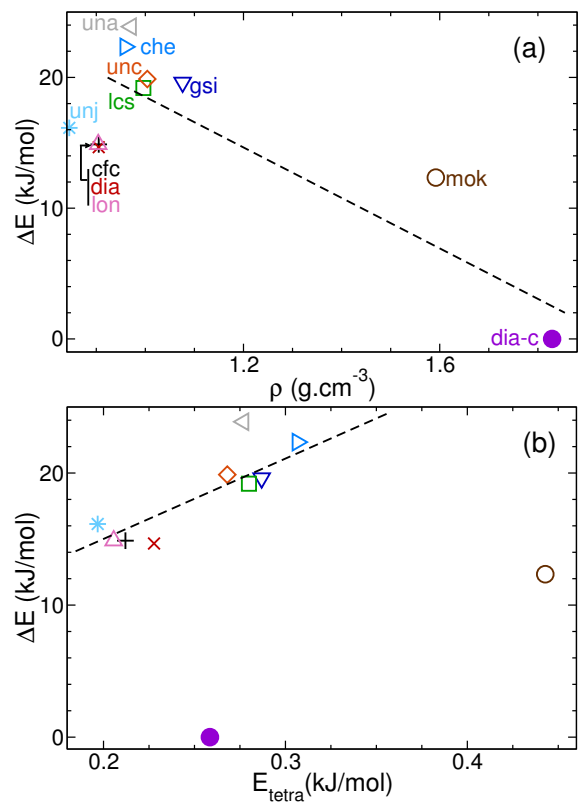

Figure 1. Relative energy of zinc cyanide polymorphs (per Zn(CN)₂ unit), using the dense *dia-c* framework as a reference, as obtained through DFT calculations. (a) Relative energy vs. density. (b) Relative energy vs. tetrahedral distortion energy (see text for details). Dashed lines are a guide to the eyes, indicating parallel linear trends.

## III. ENERGY-MINIMIZED STRUCTURES

### A. Quantum chemical calculations

We performed energy minimization at the DFT level of the 11 Zn(CN)₂ frameworks constructed from the RCSR database. Each framework relaxed to a local energy minimum. In all cases, the topology was unchanged during the minimization process: this confirms the metastability of these structures at zero Kelvin. The relative energies of the frameworks are plotted in Figure 1(a) as a function of density. The dense, interpenetrated *dia-c* framework which is the experimental structure of Zn(CN)₂ at ambient conditions, is logically found to be the most stable polymorph of Zn(CN)₂ studied. The two polymorphs obtained experimentally by fluid intrusion, *dia* and *lon*, have a similar energy of +15 kJ/mol with respect to *dia-c*, in keeping with their near identical structure. They differ mostly in the following aspect: in the *dia*, all hexagons formed of 6 neighboring vertices are in *chair-like* conformation, while in the *lon* two thirds of them are in *boat-like* conformation – in the corresponding Zn(CN)₂ frameworks, this implies differences concerning the relative orientations of neighboring coordination tetrahedra.[11]

On Figure 1(a) one can notice a correlation between density and relative energy. This generic feature is widely observed in families of inorganic polymorphs such as zeolites.[29] Several structures are found within a small energy range of the *dia-c* framework: these include the experimental structures *dia* and *lon*. The *cfc* framework is very close in energy to the latter two. One structure, the relatively dense *mok* ($\rho$ = 1.592 g.cm$^{-3}$), is lower in energy than *dia* and *lon* ($\Delta E$ = 12.5 kJ/mol), due to its relatively high density, close to that of the *dia-c* structure. Four more structures lie within 20 kJ/mol of *dia-c*, and within 5 kJ/mol of the *dia* and *lon*: *gsi*, *lcs*, *unc*, and *unj*. Given the very moderate spread in energy, these structures can thus be considered experimentally feasible. Polymorphs *che* and *una* are slightly higher, at 22 and 24 kJ/mol respectively.

| Net | cfc | che | dia | dia-c | gsi | lcs | lon | mok | una | unc | unj |
|---|---|---|---|---|---|---|---|---|---|---|---|
| Nominal density | 0.6494 | 0.6842 | 0.6495 | 1.2987 | 0.7384 | 0.6887 | 0.6494 | 1.1116 | 0.6396 | 0.6847 | 0.5989 |
| Net space group | $P6_3/mmc$ | $Ia\bar{3}d$ | $Fd\bar{3}m$ | $Pn\bar{3}m$ | $Ia\bar{3}$ | $Ia\bar{3}d$ | $P6_3/mmc$ | $Cccm$ | $P6_222$ | $P4_122$ | $P6_222$ |
| Number of vertices | 2 | 3 | 1 | 1 | 1 | 1 | 1 | 3 | 1 | 1 | 1 |
| Number of edges | 3 | 5 | 1 | 1 | 2 | 1 | 2 | 3 | 4 | 2 | 2 |
| Framework space group | $P6_3mc$ | $I2_13$ | $R3m$ | $P\bar{4}3m$ | $I2_13$ | $I\bar{4}3d$ | $P6_3mc$ | $Cc$ | $P6_1$ | $P4_1$ | $P6_1$ |
| Framework density (g.cm$^{-3}$) | 0.905 | 0.959 | 0.905 | 1.829 | 1.076 | 0.996 | 0.904 | 1.592 | 0.971 | 1.004 | 0.845 |
| Porosity | 0.352 | 0.312 | 0.352 | 0.0 | 0.235 | 0.289 | 0.349 | 0.053 | 0.312 | 0.283 | 0.397 |

Table I. List of nets studied in this work, their nominal density, space group, number of vertices and edges. For each net, we also list the space group, density and porosity of the energy-minized $Zn(CN)_2$ framework constructed from it.

Among the low-density structures studied here, it is worth noting that the lowest energy ones (*dia*, *lon* and *cfc*) share the same local connectivity and feature quasi-undistorted $[Zn(CN)_4]^{2-}$ coordination tetrahedra, while the structures with higher energy show larger dispersion in the angles around the $Zn^{2+}$ ions. In order to quantify this, we defined a simple energy term $E_{tetra}$ that quantifies the Coulomb interactions between C/N atoms around each Zn:

$$E_{tetra} = \frac{1}{24\pi\epsilon_o}\left\langle\sum_{i<j}\frac{q_iq_j}{|\vec{r}_i-\vec{r}_j|} - \frac{q_iq_j}{2d\sin(109.47°)}\right\rangle, \quad (1)$$

where the average is over all Zn atoms; for each Zn atom, in Eq. 1 $i$ and $j$ label the four coordinating C/N atoms around the central cation; the atomic charges are taken from the force field.[22]

The relative energy of the zinc cyanide polymorphs is plotted against $E_{tetra}$ on Fig. 1(b). There is a clear correlation, amongst structures of similar density (i.e. all but *mok* and *dia-c*), between tetrahedral distortion and energy. Moreover, we have checked that other factors, such as Coulombic interactions between neighboring Zn cations, play a minor role in the total energy dependence on topology. The dominant effect in the relative stability of zinc cyanide polymorphs is thus a competition between the distortion of the zinc(II) tetrahedral environment and the topology imposed by the framework, for nets whose vertices do not all have a perfect tetrahedral environment. The two exceptions are the *dia-c* structure, which is stabilized by inter-framework interactions, and the *mok* structure, which features very large distortions from the tetrahedral geometry but is also stabilized because of its higher density.

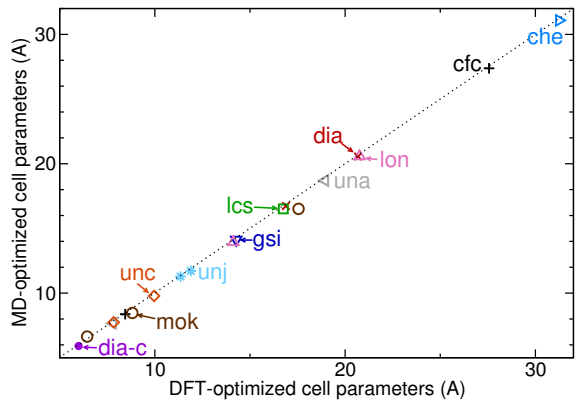

Figure 2. Comparison of DFT- with MD-optimized structures. For a given framework, each cell parameter of the MD-optimized structure, and the corresponding parameter of the DFT-optimized structure, are given on the $x$- and $y$-axis respectively.

Finally, we calculated the porosity $\phi$ of these structures based on the geometric method of Connolly,[30] i.e. the fraction of the unit cell volume that is accessible to a spherical probe of radius 1.2 Å.[41] The *dia-c* framework is nonporous, and the *mok* framework has $\phi = 0.053$. Visual inspection reveals that the porosity in *mok* is composed of small unconnected pockets (closed pores), and it thus has no accessible porosity. All other structures contain open pores, and have $0.23 < \phi < 0.40$, corresponding to estimated pore volumes of 0.253 to 0.336 g.cm$^{-3}$, and surface areas of 3550 to 3760 m$^2$.g$^{-1}$. Among these porous polymorphs, we found a very close correlation between porosity and density (see Fig. S4 in Supporting Information), owing to the very similar coordination of these frameworks.

### B. Force field validation

Given our purpose to study these frameworks at finite temperature and pressure using classical force field-based MD, one may wonder about the consistency of this approach with the previously discussed DFT calculations. That is, whether the force field validated for the *dia-c* framework of $Zn(CN)_2$ (both at ambient conditions and under pressure) is also suitable to describe other metastable polymorphs. To check that, we relaxed each polymorph wit classical MD in the limit of zero tempera-ture, and compared the resulting MD-optimized structures to the DFT-optimized ones. The agreement between structures is found to be excellent: Fig. 2 shows that cell parameters show only a tiny variation (a shortening of the order of 1 percent) from the DFT- to the MD-optimized structure. The only exception is the *mok* structure, where the variation is slightly larger ( 4%) but still quite small. Besides, Fig. S3 of the Supporting Information shows that the energy differences between these polymorphs are also well reproduced overall. The force field is thus transferable from the *dia-c* polymorph to the other structures.[42]

## IV. STABILITY AND THERMAL EXPANSION

In this Section, we describe the thermal properties of $Zn(CN)_2$ polymorphs through molecular dynamics simulations at zero pressure and in a range of temperature from 50 K to 500 K.

Molecular dynamics simulations at increasing values of temperature were used to confirm the metastability of the 11 structures generated. Such confirmation is of particular interest since it was recently shown, for a family of zeolitic imidazolate frameworks with four-connected nets similar to $Zn(CN)_2$, that a large number of the hypothetical polymorphs which had been proposed based on energy minimizations[16] were not thermally stable at room temperature.[3] While the experimental fluid-intruded structures of *dia* and *lon* polymorphs of $Zn(CN)_2$ were retained upon release to ambient pressure, and the *lon* framework appears to retain its structure upon desolvation,[11] a direct confirmation of these guest-free structures' stability is desirable. It is also unknown whether the same would be true of other potential polymorphs synthesized via fluid intrusion.

All structures were found mechanically stable at ambient temperature and indeed within the whole 50–500 K temperature range. The *dia* and *lon* polymorphs obtained experimentally, as well as the other polymorphs, should thus be amenable to fluid evacuation and should retain their porosity upon activation. This indicates the importance of molecular fluid compression as a novel way to obtain new porous structures from dense polymorphs.

We then looked at the thermal behavior of these polymorphs, especially given that the dense *dia-c*-$Zn(CN)_2$ phase is known for its very strong isotropic NTE. To compare the various polymorphs despite their different crystal symmetries, we plotted the evolutions of unit cell volume $V$ as a function of temperature in Fig. 3, and calculated for each phase the volume thermal expansion coefficient, $\alpha_V$, in the 50-500 K range:

$$\alpha_V = \frac{1}{V}\left(\frac{dV}{dT}\right) \quad (2)$$

All structures except *mok* exhibit clear NTE with a roughly linear behavior, very close low-temperature values of $\alpha_V$ (between −35 and −45 MK$^{-1}$, in reasonable agreement with the experimental value of −51 MK$^{-1}$ for *dia-c*[10,31]). Nonlinearities are mostly noticeable for dense structures: for the *dia-c*, where these non-linearities are well described by a quadratic fit, the lessening of NTE under heating is comparable to the one measured in Ref. 32: $\alpha_V \simeq -44$ MK$^{-1}$ for $T \to 0$, while $\alpha_V \simeq -24$ MK$^{-1}$ near 400 K. An important feature is that this contraction under heating is essentially isotropic with linear expansion coefficients $\alpha_{a/b/c}$ very close to each other (see Table I in Supporting Information).

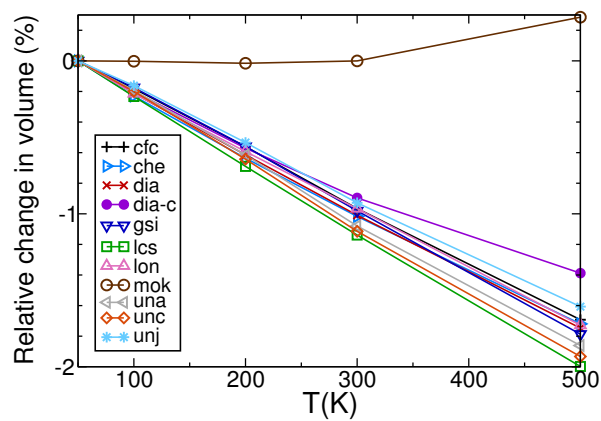

Figure 3. Relative volume change (w.r.t. the value at 50$K$) versus temperature, for the 11 structures of this study.

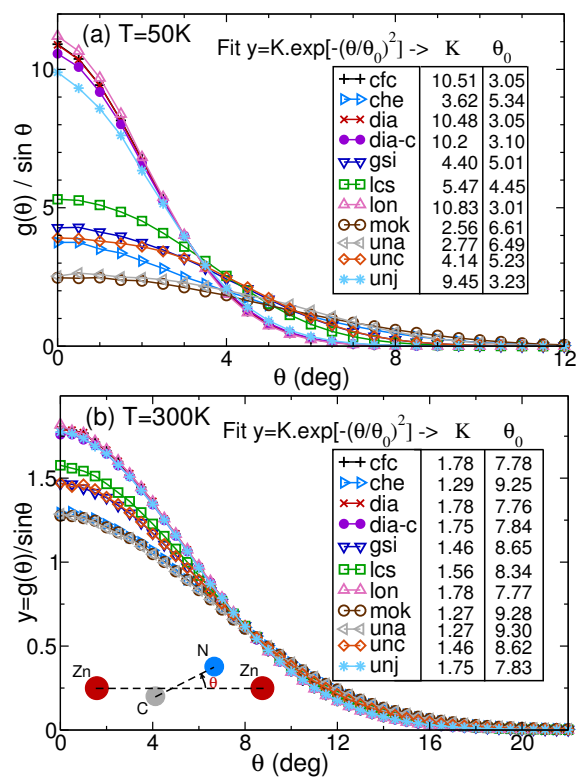

Figure 4. Functions $g(\theta)/\sin\theta$ where $g(\theta)$ is the distribution of angle $\theta$ [see inset of (b)], computed for the structures listed, at $T = 50$ K (a) and $T = 300$ K (b). Gaussian fits were done, with parameters $K$ and $\theta_0$ indicated in the table.

This near-universality of NTE (and its extent) among the zinc cyanide frameworks, coupled with its isotropic character and its occurence even in frameworks derived from non-bipartite nets such as *unj*, indicates that the mechanism at play for NTE does not rely necessarily on rotations of pairs of tetrahedra (involving off-centering of the connecting cyanide)[33] but can also be compatible with correlated motions of larger units. Consistently, the isotropic character of NTE indicates that it relies on a local, topology-independent mechanism, that may well be buckling vibrations of the Zn–C–N–Zn linkages (as long as tetrahedron distortions are small enough to allow for such vibrations). We provide here an analysis of the buckling vibrations of the Zn–C–N–Zn linkages, whose importance is clearly demonstrated by the distribution $g(\theta)$ of the angle $\theta$ between a given cyanide axis and the associated Zn–Zn axis (see Fig. 4). We computed those from the atomic trajectory data of MD simulations at $T = 50$ K and $T = 300$ K. Fits of these distributions indicate clearly a behavior of the form

$$g(\theta) = K \sin \theta e^{-(\theta/\theta_o)^2} \qquad (3)$$

where $K$ is a normalisation constant such that $\int_0^{\frac{\pi}{2}} g(\theta)d\theta = 1$ – the $\sin \theta$ prefactor stands for the solid angle associated with an interval $[\theta; \theta + d\theta]$. Whereas the typical angular extension ($\theta_o$) doesn't depend much on the framework at $T = 300$ K, at lower temperature one can distinguish several groups: (i) structures like *dia*, *lon* or *unj* where $\theta$ in the $T = 0$ K structures is negligible ($\leq 1$ degree), and $\theta_o$ is roughly proportional to $\sqrt{T}$; (iii) structures like *mok* or *una*, with large tetrahedron distortion and $\theta$ values exceeding 5 degrees on some bonds of the $T = 0$ structure, which significantly enhance $\theta_o$ especially at low temperatures; (ii) intermediate cases. These distributions evidence that, even at $T = 50$ K, kinetic energy is sufficient for cyanide ions to escape the shallow potential wells resulting from mutually frustrated force fields, and to vibrate around the Zn-Zn axis. Assuming that these vibrations can combine efficiently to form, at a global scale, Rigid Unit Modes (RUM) at low energy (which is the case except for the *mok*), they allow for a large negative $\alpha_V$ value.

The *mok* structure is here again a particular case: its thermal expansion coefficient is nearly zero. Indeed, distortions of the coordination tetrahedra in the relaxed structure are so large that they hinder tetrahedron rotations and push to higher energy the associated RUMs that would account for NTE otherwise.

## V. MECHANICAL BEHAVIOR

### A. Stability under pressure

After checking that our predicted polymorphs were stable under ambient conditions ($T = 300$ K, $P = 0$), we now address the questions of mechanical stability and behavior under pressure. To this aim, simulations with increasing values of pressure were carried out in a sequential manner, starting from $P = 0$ and increasing pressure stepwise by $\Delta P = 0.05$ GPa. Results of these *in silico* compression experiments are presented on Fig. 5, under the form of a volume vs pressure plot. All porous phases have limited stability, undergoing a discontinuous transition at moderate critical pressures ranging from $P_c \simeq 0.2$ GPa (*unc*) to $P_c \simeq 0.5$ GPa (*dia*, *lon* and *cfc*). The case of the *gsi* structure stands out somewhat from the others: data up to 0.4 GPa indicate a possible continuous transition near $P \simeq 0.2$ GPa, but no associated symmetry breaking was evidenced; in any case, the structure was found to be unstable at $P \geq 0.45$ GPa. Comparison of $P_c$ values between different structures indicates that $P_c$ clearly increases with the density, and for porous structures decreases somewhat with increasing tetrahedral distortions (see Fig. S6 in Supporting Information).

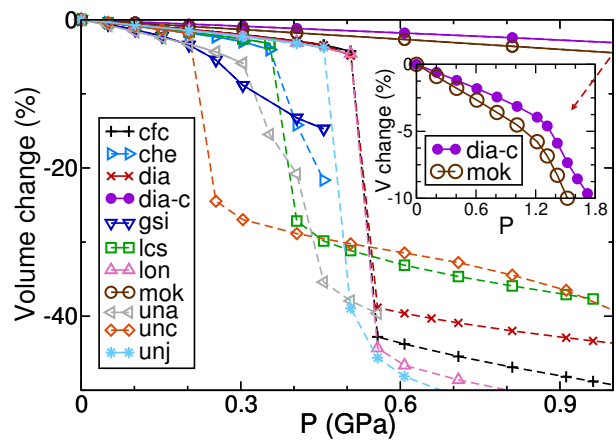

Figure 5. Relative volume change, w.r.t. the zero pressure value, at $T = 300$ K and for the 11 structures listed of Zn(CN)$_2$.

This sub-gigapascal limit of mechanical stability of porous structures contrasts with the behavior of the denser *dia-c* and *mok* frameworks: both are stable to pressures exceeding 1 GPa, following a generic density/mechanical stability correlation.[3] The limit of stability of the *dia-c* structure is found at $P_c \simeq 2.1$ GPa. This behavior was already reported in Ref. 22, and put in parallel with the discontinuous transition towards a high-pressure phase, observed experimentally around the same pressure.[34] Additionally, for both dense structures, a kink in the $V(P)$ curve is found at $P'_c \simeq 1.2$ GPa, reflecting the experimentally-observed transition to a high-pressure phase (labelled ''phase II'' in ref. 34). The *mok* structure shows the same behavior as *dia-c* under compression. From the analysis of the *dia-c* and *mok* trajectories, we can show that this high-pressure transformation involves a shortening of the distances between a zinc and its nearest non-coordinating cyanides (see Fig. S5 of the Supporting Information, and accompanying text), hinting to a move towards higher coordination around the zinc, as evidenced experimentally by Collings et al.[34]

The behavior of porous structures under compression can also be analyzed by considering the distributions of the cyanide's buckling angle, $\theta$ (see Fig. 4). In the low-pressure phases (for $P < P_c$) the distributions $g(\theta)$, shown in Figs. S8,S9,S10 of the Supporting Information, still follow the law given in Eq. 3. The typical angular extension $\theta_o$ increases with pressure, as increasing $\theta$ allows to decrease cell parameters without decreasing the distances between Zn and its coordinating atoms. In the high pressure phases of porous structures, this type of distributions is also observed, however the $\theta_o(P)$ plots are characterized by a jump at $P_c$. In fact, the transition is better characterized by (i) the distribution $g(d)$ of the *distance* $d$ between a given cyanide ion and the associated Zn–Zn axis (see inset of Fig.6-c), or (ii) the radial distributions $g(r_{Zn-X})$ (either $X = Zn$ for Zn-Zn distances or $X = C/N$ for distance between Zn and either C or N).

On Fig. 6, for the porous structures *dia* and *lcs*, one can clearly distinguish the unimodal distributions $g(d)$ in the low-pressure phase from those in the high-pressure phase, which are bimodal, indicating the coexistence in this phase of two types of Zn–CN–Zn linkages, one of them being much more distorted than the other. The case of the dense *dia-c* is different, and consistent with a transition of a continuous nature. At $P'_c \simeq 1.2$ GPa, the distribution remains monomodal but one notices a clear change of behavior: the distribution extension (or $\langle d \rangle$) is almost unaffected by $P$ in the low-pressure regime, while for $P > P'_c$ it increases much faster. Although $\langle d \rangle$ cannot strictly be considered as an order of parameter for this transition (it is non-zero in the low-pressure phase), its behavior supports the picture of a continuous transition induced by a specific phonon mode softening, and involving cyanide ions off-centering drifting away from their respective Zn–Zn axes. This conclusion can also be reached by the analysis of the distributions of radial distances $g(r_{Zn-C/N})$ (see Fig. S5 of the Supporting Information, and accompanying text).

### B. Response in the low pressure regime

We now turn our attention to the mechanical properties of the structures predicted in this work. For each framework, we performed a fit of the volume–pressure curve by a third-order Birch-Murnaghan equation of state[35] in the low-pressure region:

$$P(V) = \frac{3B_0}{2}\left[\left(\frac{V_0}{V}\right)^{\frac{7}{3}} - \left(\frac{V_0}{V}\right)^{\frac{5}{3}}\right]$$
$$\times \left[1 + \frac{3}{4}(B'_0 - 4)\left(\left(\frac{V_0}{V}\right)^{\frac{2}{3}} - 1\right)\right] \quad (4)$$

to obtain zero-pressure values of bulk modulus, $B_0 = B(P = 0)$, and of its derivative with respect to pressure, $B'_0 = \frac{dB}{dP}(P = 0)$. Bulk moduli show much dispersion among the 11 structures considered [Fig. 7]. The largest ($B_0 = 34.6$ GPa) is found for the densest, *dia-c* structure. In fact, we observe a good correlation between $B_0$ and other structural factors, in particular the density $\rho$ and the relative energy $\Delta E$. Moreover, among the porous polymorphs, we find a close correlation between the bulk modulus and the amplitude of buckling modes $\theta_0$ (at 300 K). Finally, we also observe that the stiffest structures (high $B_0$) feature higher resistance to pressure (See Fig. S6(c) in Supporting Information).

All polymorphs studied here have a negative value of $B'_0$ (see Fig. S7 in Supporting Information), reflecting an anomalous decrease in bulk modulus with increasing pressure, or pressure-induced softening.[36] This behavior, first demonstrated in the

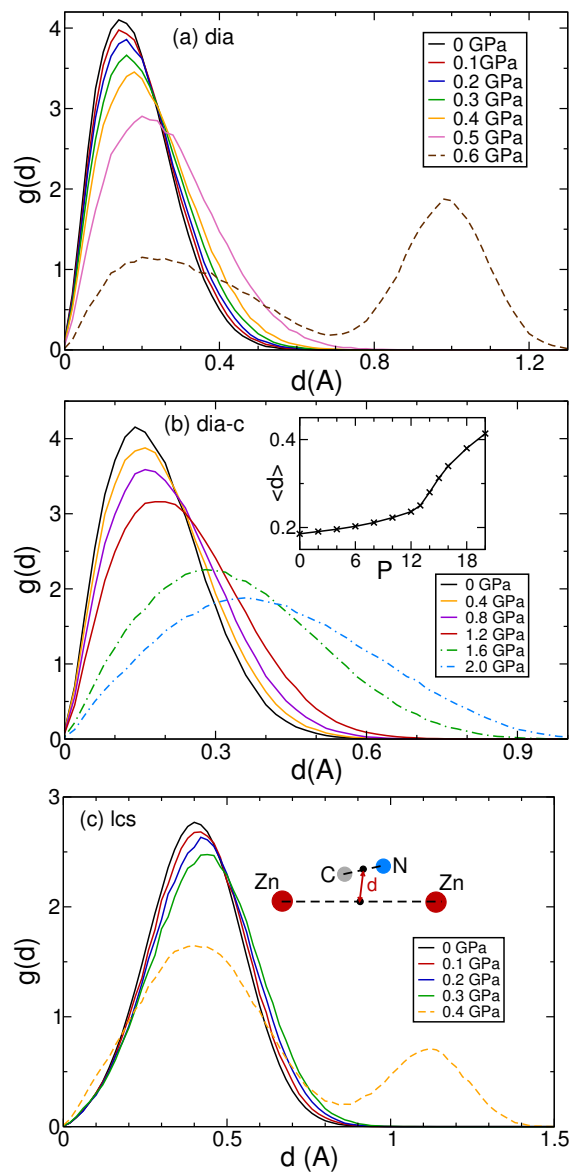

Figure 6. Distributions of the distance $d$ (in ångströms) between (the center of mass of) a CN ion and the middle of the associated Zn–Zn linkage, for the *dia* (up-left), *lcs* (up-right) and *dia-c* (down) structures. $g(d)$ was computed from atomic trajectory data of MD simulations at $T = 300$ K, and pressures $P$ indicated in the captions. For *dia-c*, the inset shows the evolution of the average distance $\langle d \rangle$ with pressure.

*dia-c* phase in Ref.[10] ($B'_0 = -6.0(7)$) and later measured in various pressure-transmitting media ($B'_0 = -8.6(4)$ in Ref. 34; $B'_0 = -8.48(15)$ in Ref. 11), is apparently a universal feature of $Zn(CN)_2$ polymorphs. Our force field-based simulations show only qualitative agreement with those experimental measurements, with a value of $B'_0 = -4.0$ for *dia-c*. Nevertheless, this allows us to compare the extent of softening between the structures studied. We find that the value of $B'_0$ obtained for the porous polymorphs is much larger than that of *dia-c*, with values around −10 for the *dia*, *lon*, *cfc* and *unc* phases, and −16 for *gsi*. This is particularly large compared to known miner-als exhibiting pressure-induced softening, such as malayaite ($B'_0 = -3$). From comparing values of $B'_0$ in different structures, it seems that $B'_0$ is less negative for dense and/or highly distorted structures (*dia-c*, *mok*, *una*); but no simple relationship could be established between $B'_0$ and any of the frameworks' physical properties (density, relative energy, etc.). While the existence of universal pressure-induced softening in the Zn(CN)$_2$ family is inherent to the chemistry of the Zn–C–N–Zn linkages, and linked to its universally-shared anomalous thermal expansion, its extent is dependent on the structure's symmetries, via the nature and repartition in **k**-space of rigid unit modes accounting for it.[22,37]

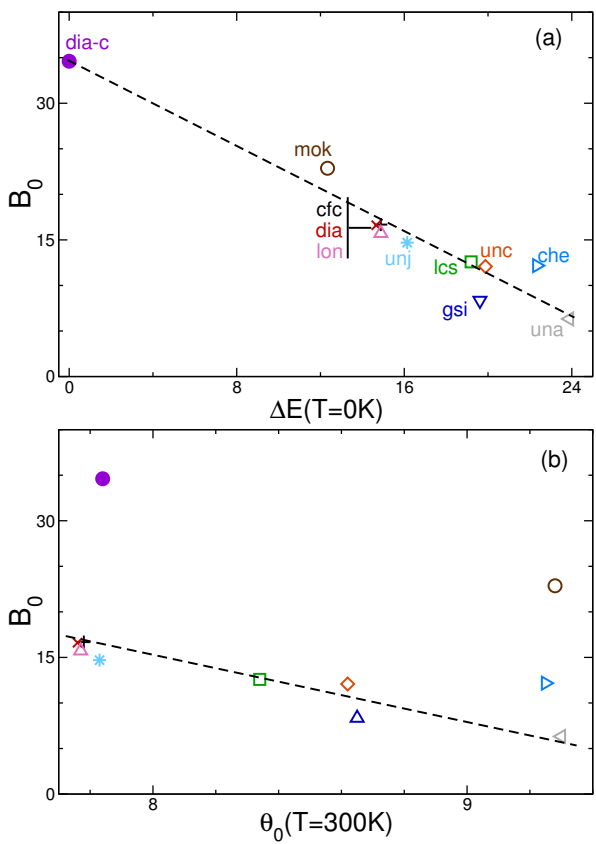

| Phase | $\beta_{min}$ | $\beta_{max}$ | $E_{min}$ | $E_{max}$ | $G_{min}$ | $G_{max}$ | $\nu_{min}$ | $\nu_{max}$ |
|---|---|---|---|---|---|---|---|---|
| *cfc* | 19.6 | 22.4 | 16.0 | 25.2 | 1.6 | 13.2 | −0.32 | 1.02 |
| *che* | 26.3 | 28.6 | 6.2 | 34.7 | 2.17 | 17.4 | −0.53 | 1.09 |
| *dia* | 19.1 | 22.9 | 11.1 | 30.9 | 3.5 | 17.1 | −0.22 | 0.86 |
| *dia-c* | 9.3 | 10.0 | 9.1 | 83.3 | 3.6 | 50.2 | −0.65 | 1.31 |
| *gsi* | 37.7 | 40.4 | 8.5 | 27.4 | 3.3 | 15.5 | −0.40 | 0.73 |
| *lcs* | 24.6 | 26.9 | 8.6 | 21.9 | 3.2 | 10.6 | −0.18 | 0.75 |
| *lon* | 18.5 | 28.0 | 15.0 | 24.8 | 1.2 | 13.1 | −0.51 | 1.24 |
| *mok* | 11.3 | 20.8 | 18.8 | 55.6 | 4.6 | 29.3 | −0.32 | 0.95 |
| *una* | −5.7 | 10.4 | 8.4 | 23.2 | 0.60 | 18.7 | −0.96 | 1.63 |
| *unc* | 6.7 | 40.6 | 9.3 | 23.1 | 2.8 | 20.3 | −0.41 | 1.05 |
| *unj* | 19.1 | 34.7 | 7.1 | 26.1 | 0.68 | 13.8 | −0.81 | 1.53 |

Table II. Minimal and maximal values, for each structure at $T = 300K$, of: linear compressibility $\beta$ (in TPa$^{-1}$), Young's modulus $E$ (in GPa), shear modulus $G$ (in GPa) and Poisson's ratio.

Figure 7. Bulk moduli $B_0$ (in GPa) estimated from MD simulations at $T = 300K$ and subsequent fits to Eq. 4, for the 11 structures indicated; they are plotted against (a) the DFT-obtained energy $\Delta E(T = 0K)$ (in kJ/mol); and (b) the angle $\theta_0$ (in degrees).

## C. Mechanical properties

There has been a lot of interest recently in the search for framework materials with anomalous mechanical properties, including negative linear compressibility,[4,5] negative Poisson's ratio[7,8] and highly anisotropic mechanical behavior.[5] We have thus further quantified the mechanical response of the proposed Zn(CN)$_2$ polymorphs by carrying out long MD runs (total time: 5 ns) at $T = 300$ K and $P = 0$ for each of the 11 structures. From the fluctuations of cell parameters, their stiffness tensors $C_{ij}$ ($1 \le i, j \le 6$ in Voigt notation) can be extracted:

$$C_{ij}^{-1} = \left( \frac{V}{k_B T} \right) \left( \langle \epsilon_i \epsilon_j \rangle - \langle \epsilon_i \rangle \langle \epsilon_j \rangle \right) \qquad (5)$$

(with $\epsilon$ the strain tensor). These in turn give access to various direction-dependent quantities, such as the linear compressibility $\beta(\mathbf{u})$, the Young's modulus $E(\mathbf{u})$, or the shear modulus $G(\mathbf{u}, \mathbf{v})$ which quantify respectively the deformation along $\mathbf{u}$ subsequent to a uniform compression, the deformation along $\mathbf{u}$ following a compression along this axis, and the resistance of the plane $\mathcal{P}$ normal to $\mathbf{v}$ to shearing along $\mathbf{u} \in \mathcal{P}$.[5] We used the ELATE online tool to perform this analysis.[38]

Concerning the linear compressibility, only one structure exhibits negative linear compressibility: the *una*, which expands slightly along one direction (the *c* axis) upon isostatic compression: $\beta(\mathbf{c}) \simeq -5$ TPa$^{-1}$. The *c* axis in this structure is that of porous channels; unlike the *wine-rack* structure of some Metal-Organic Frameworks[39], where 1D porous channels along a given axis allow for a large NLC in a direction transverse to channels, here the 6-fold symmetry of the channel lattice disables this effect. Yet, this small NLC along *c* is reminiscent of the very small compressibility along the channel direction in the *wine-rack* case. Other structures with well-defined linear channels have also large anisotropy in linear compressibility, e.g. the *unc* has $\beta(\mathbf{c}) = 6.7$ TPa$^{-1} \simeq \beta_{in-plane}/6$. In contrast, structures for which the framework is of cubic symmetry, or at least contains distinct types of channels with non-collinear axes, have much smaller anisotropy in linear compressibility, $\beta_{max}/\beta_{min} \le 1.2$.

All structures show substantial anisotropy in Young's moduli. The anisotropy ratio $E_{max}/E_{min}$ varies from values $\simeq 1.6$ (*cfc* and *lon* structures) to 9.1 for the *dia-c*. The latter value can appear large for a highly symmetric and dense structure. In fact, there, the Young's modulus is maximized along the [111] direction (83.3 GPa), and those equivalent by symmetry; compression along such an axis, due to the interpenetration of frameworks would necessarily involve compression of Zn-C-N-Zn linkages along this axis. Thus, the material is much stiffer along these directions than along crystallographic axes like [100], along which compression can occur by e.g. deforming coordination tetra-hedra and without compressing individual bonds. For porous structures, where void space allows more easily for compression by angular distortions, the maximal Young's modulus is much smaller (reaching 34.7 GPa somehow for the *che* structure). The minimal Young's modulus can be as small as 6.2 GPa and 7.1 GPa for the *che* and *unj* respectively. In the latter case it corresponds to a compression along 1D channels, which is probably easier since the walls of these channels are made of edge-sharing pentagons (each edge consisting of a Zn-C-N-Zn linkage, and none of these edges is along the channel axis), and thus relatively easy to deform. The *che* case, though with a more complex structure, also allows easy compression along certain axes, forming substantial angles ≥ 30 degrees with all Zn–C–N–Zn linkages of the framework.

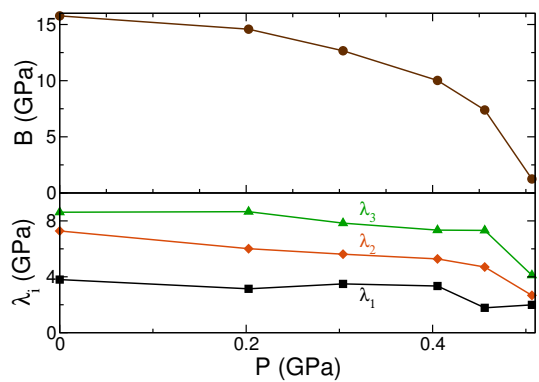

Figure 8. Mechanical properties of the *dia* framework versus pressure, studied by 10 ns-long MD runs, at $T$ = 300 K. Up: Bulk modulus $B(P)$; Down: Three lowest eigenvalues $\lambda_i$ of the $6 \times 6$ stiffness tensor.

The response to uniaxial stress is also characterized by the Poisson's ratio $v(\mathbf{u},\mathbf{v})$ (ratio of transverse strain along $\mathbf{v}$ to strain along the compression axis $\mathbf{u}$). Remarkably, we find this ratio to be negative in some directions for all the structures studied. The most negative values are found (i) for structures with well-defined 1D channels, like *una* or *unj*; (ii) the *dia-c* structure. Auxeticity in these materials seem to result from a mechanism already encountered in zeolite frameworks, involving correlated rotations of neighboring tetrahedra.[8]

Finally, in order to confirm the nature of the pressure-induced phase transition in porous phases of $Zn(CN)_2$, we performed calculations of elastic constants of the *dia* structure at increasing values of pressure up to $P_c$. The bulk moduli, $B(P)$, calculated at each pressure from the elastic tensors, is shown in Fig. 8. It can be seen to decrease with pressure, following a power law ~ $(P_c - P)^\alpha$, with exponent $\alpha \simeq 0.36$. Moreover, when approaching the transition, the lowest eigenvalues of the stiffness matrix (see Fig. 8) decrease significantly; an analysis of the associated eigenvectors gives indication for a shearing mode, stimulated by compression along a transverse axis (i.e. non-collinear to cell vectors $\mathbf{T}_{a/b/c}$). This demonstrates that the instability under pressure is due to pressure-induced shear mode softening, similar to that observed for several porous metal-organic frameworks.[3,40]

## VI. CONCLUSIONS

Following the recent report that $Zn(CN)_2$ can undergo reconstructive transitions to porous polymorphs upon compression in molecular fluids, we have investigated the feasibility of ten different four-connected nets as hypothetical zinc cyanide polymorphs. Through a combination of quantum chemical calculations and molecular dynamics simulations, we have confirmed the metastability of the two phases recently evidenced experimentally, and suggest the existence of 7 novel porous phases of $Zn(CN)_2$ with relatively small formation enthalpy from the dense polymorph. Furthermore, we have characterized their thermal and mechanical behavior, showing that isotropic negative thermal expansion is a near-universal feature of this family of materials, with thermal expansion coefficients somewhat larger than that of the dense phase. In contrast, we find a wide variety in the mechanical behavior of these porous structures, with framework-dependent anisotropic compressibilities. All porous structures, however, show pressure-induced softening leading to a structural transition at modest pressure.

### ACKNOWLEDGEMENTS

We thank Aurélie Ortiz, Alain Fuchs, and Andrew Goodwin for helpful comments and discussions. We acknowledge the access to HPC resources from GENCI (grant x2015087069).

### SUPPORTING INFORMATION AVAILABLE

DFT-optimized structures; input files for DFT and molecular dynamics calculations; representations of the frameworks studied; table of linear and volumetric thermal expansion coefficients; values of minimal and maximal Zn–Zn–Zn angles; plots showing energy and porosity versus density (at $T$ = 0 K and $P$ = 0 GPa); distributions of distances for various pressures in 2 frameworks; plots showing the critical pressure $P_c$ and the coefficient $B'_0$ at $T$ = 300 K; distributions of angles $\theta$ at various pressures and evolution of cell parameters versus $P$ (at $T$ = 300 K). This information is available free of charge via the Internet at http://pubs.acs.org.

\* Email: anne.boutin@ens.fr

† Email: fx.coudert@chimie-paristech.fr.
 Twitter: @fxcoudert.
 Web: http://coudert.name/